\documentclass[
    ,final            
  ]
  {aipproc}
\usepackage{epsfig}

\layoutstyle{6x9}

\begin{document}

\title{Neutral Current $\pi^0$ Production in MiniBooNE}


\author{J.M. Link\footnote{For the MiniBooNE Collaboration.}}{
  address={Virginia Polytechnic Institute and State University, Blacksburg, VA}
}

\begin{abstract}
This paper describes the analysis used to determine the neutral current $\pi^0$ production in MiniBooNE in bins of momentum.  Additionally, a measurement of the relative coherent production of $\pi^0$s is discussed.  The coherent production rate is found to be (19.5 $\pm$1.1 (stat) $\pm$2.5 (sys))\% of the total exclusive neutral current $\pi^0$ production rate.
\end{abstract}

\maketitle


\section{Introduction}
Neutral current $\pi^0-$ production is a potential major background to the $\nu_e$ appearance signal that MiniBooNE is looking for in its test of LSND~\cite{Aguilar:2001ty}.  As such the main objective of this analysis is to measure the rate of $\pi^0$ production so that misidentification in the $\nu_e$ oscillation sample can be determined as a function of reconstructed $\nu_e$ energy in the charged current quasi-elastic mode.  As a result the initial product of the NC $\pi^0$ analysis effort is not an absolute cross section.  Instead it is a measurement of the total $\pi^0$ production in bins of momentum, and a measurement of the coherent production which effectively fixes the angular distribution.  The analysis described here was used in the recent oscillation analysis reported by MiniBooNE~\cite{AguilarArevalo:2007it} to fix the $\pi^0$ production in the Monte Carlo based on the observed $\pi^0$ rates in data.  In addition, the dynamics of neutrino induced, neutral current $\pi^0$ production is of interest in its own right.  The analysis shown here builds upon the work previously shown at NuInt04~\cite{Raaf:2004ty}.

\section{Determining $\pi^0$ Production in Momentum}
The event selection begins with a set of pre-cuts that exactly match the pre-cuts used in the electron neutrino selection.  The event must have only a primary event without evidence of a secondary event consistent with a muon decay  (or Michel electron).  This eliminates the vast majority of charged current $\nu_{\mu}$ interactions.  The event must have more than 200 hits in the main tank.  This is well above the Michel endpoint.  The event must have fewer than 6 veto hits.  This eliminates more that 99.9\% of all cosmic rays.  Additionally, all events must be in the 1.6~$\mu$s beam spill window, although by the time the all other cuts are applied this is essentially all that remains.  The determined production rates are all relative to these pre-cuts.  Therefore, if one is interested in computing a cross section from these numbers, it would be important to understand the inefficiency of these cuts (for example from the overlap of two neutrino interactions, or of a single neutrino interaction with a cosmic ray) and the effective target volume, which is largely set by the veto cut.  

The analysis cuts are based on the reconstruction which fits each event with  muon, electron and $\pi^0$ hypotheses.  Each fit produces a likelihood, and the log of the ratio of different likelihood hypotheses are used for particle identification.  In the first stage, electron-{\it like} events are selected by applying the cut $\log(\mathcal{L}_{\mu}/\mathcal{L}_e)>0.05$\@.  Next the  $\pi^0$-{\it like} events are selected with $\log(\mathcal{L}_{\pi}/\mathcal{L}_e)<0$\@.  When coupled with a reconstructed $\gamma\gamma$ mass cut, about the $\pi^0$ mass, a very clean sample of $\pi^0$ events (signal to noise ratio $\sim$30) is selected with an overall efficiency of about 40\%.

\begin{figure}[b!]
\includegraphics[width=14.5cm]{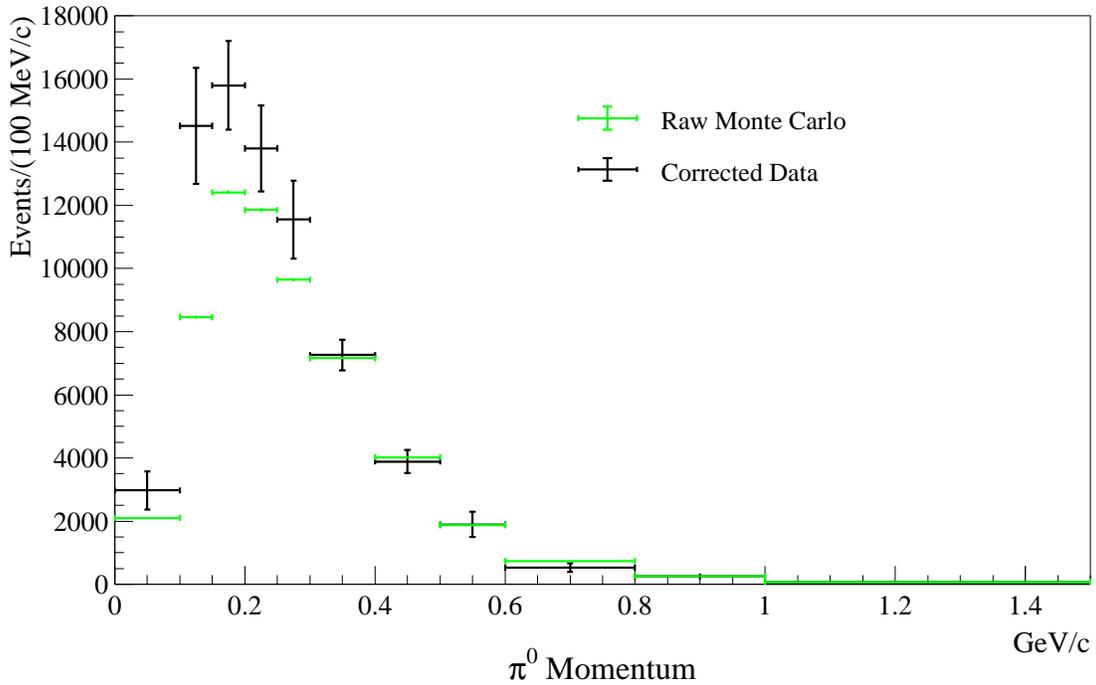}
\caption{\label{mom_corr1} Results of the $\pi^0$ unsmearing in bins of momentum. The blue points show the corrected pi0 momentum distribution and the red points show the raw Monte Carlo $\pi^0$ momentum distribution.  This comparison is shown absolutely normalized.}
\end{figure}
\begin{figure}[tbp]
\includegraphics[width=14.5cm]{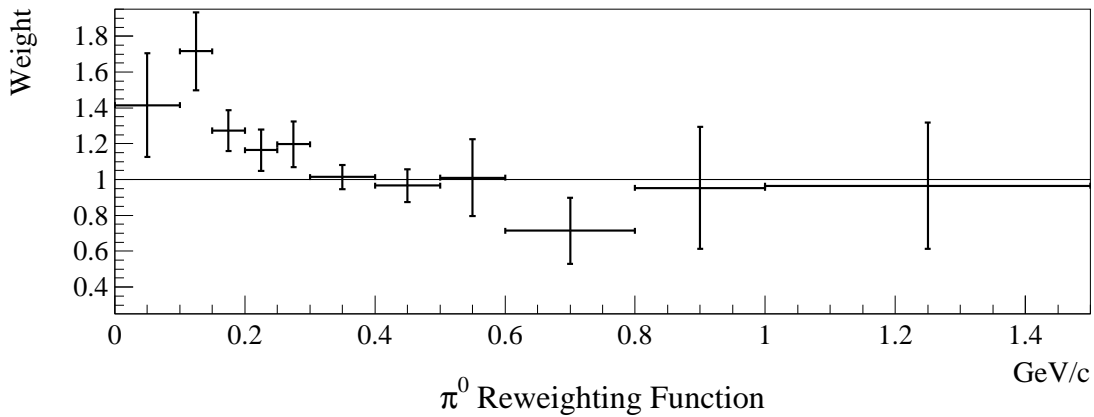}
\caption{\label{mom_corr2} The $\pi^0$ reweighting function, which is used to correct the $\pi^0$ rate of the Monte Carlo in bins of momentum.  This function is just the ratio of the distributions shown in Figure~\ref{mom_corr1}.}
\end{figure}
\begin{figure}[tbp]
\includegraphics[width=12.5cm]{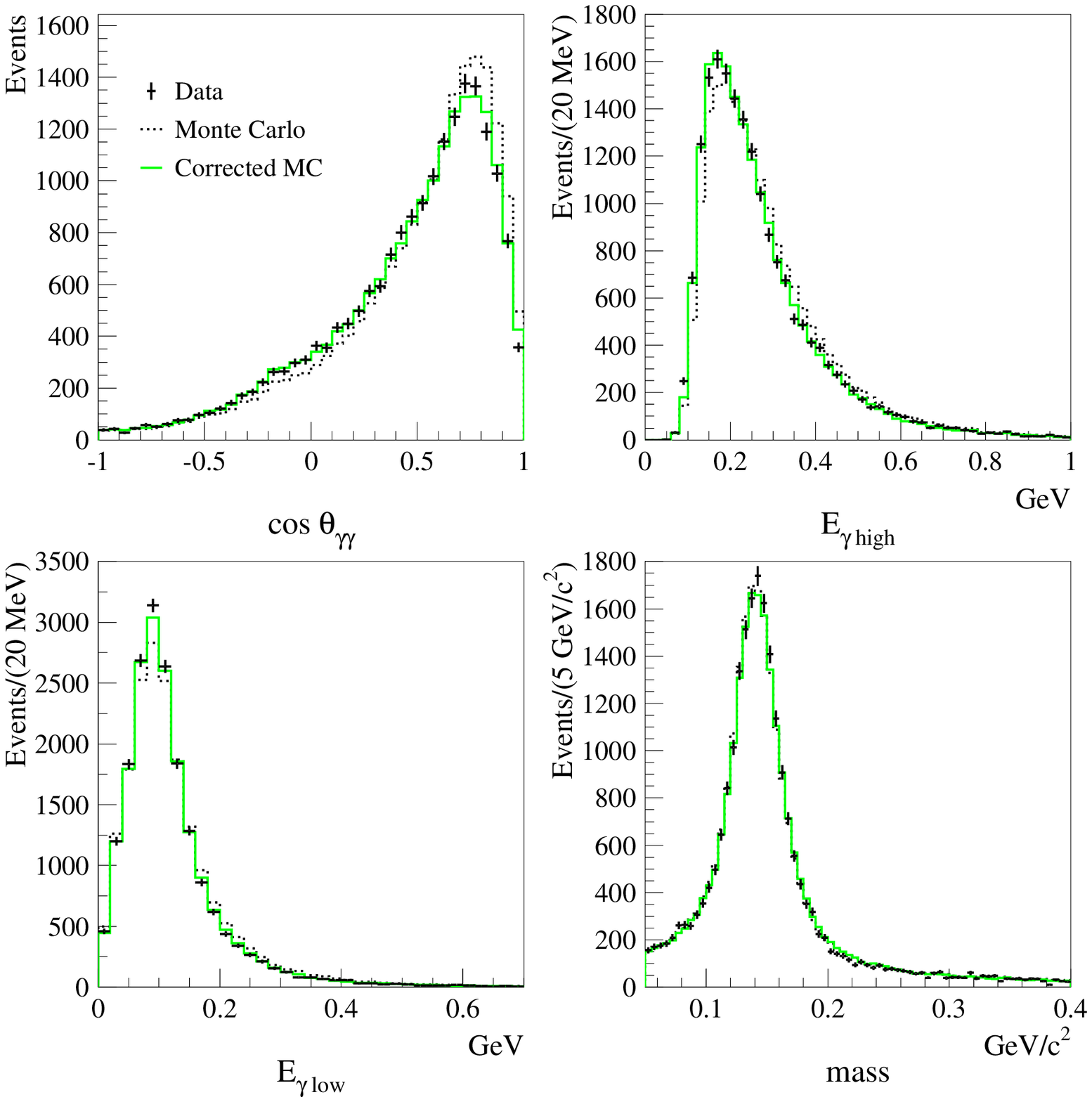}
\begin{picture}(0,0)
\put(-320,340){a)}
\put(-140,340){b)}
\put(-320,162){c)}
\put(-140,162){d)}
\end{picture}
\caption{\label{mom_corr3} A comparison of raw and corrected Monte Carlo to data for various $\pi^0$ kinematic distributions: a) the opening angel between the two gammas, b) energy of the most energetic gamma, c) energy of the least energetic gamma, d) $\gamma\gamma$ mass. In all plots the corrected Monte Carlo is in as good or better agreement with the data than is the raw Monte Carlo.}
\end{figure}

The $\pi^0$ candidate events are divided into bins of reconstructed $\pi^0$ momentum, and the Monte Carlo (MC) is used to unsmear the data, correcting for momentum smearing and inefficiency.  A matrix is formed by dividing MC events into bins of true momentum verse reconstructed momentum and counting events over background in each bin.  Each matrix element is divided by the total number of events in all reconstructed bins with the same true momentum range.  This matrix, which is well conditioned and largely diagonal, is inverted to form the unsmearing matrix.  The data vector is formed by dividing the candidate events into the same reconstructed bins and subtracting the background in each bin according to the signal to noise ratio for the corresponding MC reconstructed range.  The unsmeared, or corrected, data rates are the product of the unsmearing matrix and the data vector.  Figure~\ref{mom_corr1} shows an absolutely normalized comparison of the raw MC prediction to the corrected data distribution.  The ratio of these two distribution forms a correction function (Figure~\ref{mom_corr2}) which is used to reweight $\pi^0$ events as a function of true momentum, in the MC.   

By construction, the reweighting fixes the discrepancy between data and MC in reconstructed $\pi^0$ momentum.  Additionally, it also improves agreement in many of the key kinematic distributions.  Figure~\ref{mom_corr3} shows the relatively normalized data to MC comparison for both the raw and corrected MC.  The kinematic distributions shown are the cosine of the $\gamma\gamma$ opening angle, the $\gamma$ energies and the $\gamma\gamma$ mass. The opening angle and energy comparisons show marked improvement, while agreement in the mass distribution is largely unaffected by the reweighting.  

\section{Resonant and Coherent $\pi^0$ Production}
In neutrino-nucleus interactions, there are two main mechanisms for $\pi^0$ production.  The $\pi^0$ can result from the decay of a resonance, such as a $\Delta^+$ or $\Delta^0$, that was produced in the primary interaction, or it can be produced coherently off of the entire nucleus.  Coherent and resonant production have very different distributions for the pion angle with respect to the beam direction -- the extra mass of the resonance tends to broaden the angular distribution, while the coherent pions pile-up in the forward direction.  This fact can be used to fit the relative contribution of the two production mechanisms.

\begin{figure}[tbp]
\includegraphics[width=14.5cm]{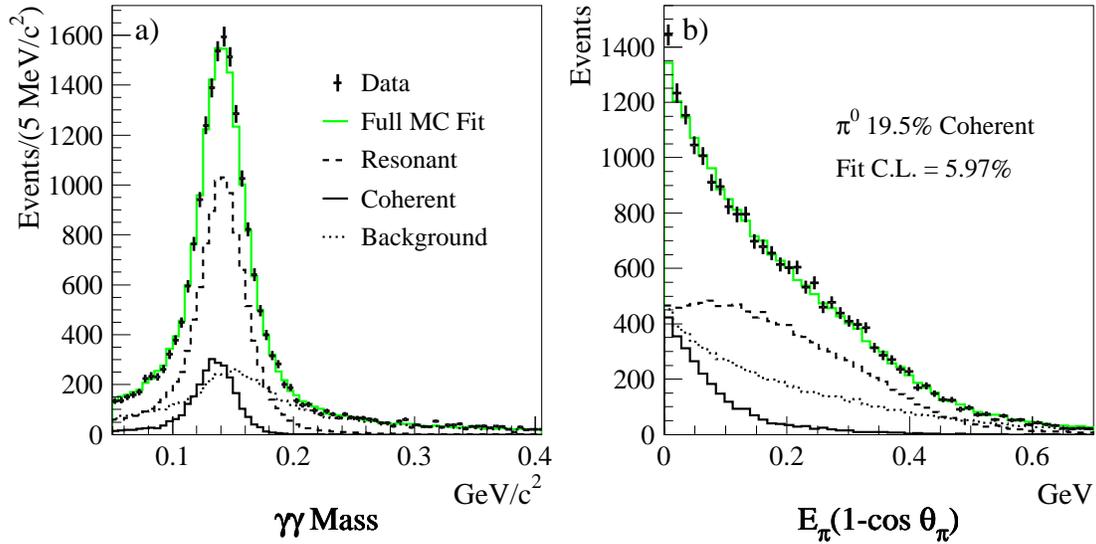}
\begin{picture}(0,0)
\put(-367,192){a)}
\put(-160,192){b)}
\end{picture}
\caption{\label{coh_fit} Monte Carlo overlay of the template fitted $\pi^0$ data in a) $\gamma\gamma$ mass, and b) $E(1-\cos\theta)$.}
\end{figure}

The $\pi^0$ candidate events in the momentum reweighted MC are used to form three templates: one for resonant events, one for coherent events, and one for background.  While the angular distribution for the coherent and resonant events are quite different, the coherent and background events are somewhat similar.  So the templates are made in two dimensions: the first dimension is a function of angle ($E_{\pi}(1-\cos\theta)$) and the second dimension is mass.  The more complex angular function is used because it has a consistent shape for coherent events across all $\pi^0$ momenta at MiniBooNE energies.  Variable binning is used such that the total number of MC events in each bin is approximately equal.  The number of bins in each projection is varied, independently, from 15 to 25, for a total of 121 fits, and the average fit parameters are used.  Figure~\ref{coh_fit} shows the fit result plotted against data in the two projections.  For the MiniBooNE flux and detector, with the NUANCE generator~\cite{Casper:2002sd} providing the secondary interaction model, the fit finds that (19.5$\pm$1.1)\% of all exclusive neutral current $\pi^0$ production is coherent.  This should be compared to the raw MC which predicts 30\% coherent for the MiniBooNE flux and detector. 

To determine the best overall production parameterization, the binned momentum and coherent fits are iterated.  The iteration converges after only one round.  In the oscillation analysis, the MC is corrected in both momentum and coherent fraction.

In the coherent analysis, a number of possible sources of systematic error were investigated including: choice of binning, background composition, momentum reweighting, neutrino flux, choice of analysis cuts, and detector model.  Table~\ref{coh_sys} lists the error estimation from each of these sources.  By far, the largest source of systematic error is the detector model, which comes primarily from the uncertainty in the reconstructed energy scale.

\begin{table}[b]
\begin{tabular}{@{\extracolsep{3.0cm}}lc}
\hline
\tablehead{1}{l}{b}{Source}  &
\tablehead{1}{c}{b}{Error (\%)} \\ 
\hline 
Binning          & 0.21\\
Background Model & 0.64\\
Reweighting      & 0.51\\
Flux             & 0.06\\
Analysis Cuts    & 0.51\\
Detector Model   & 2.34\\
\hline
All Systematics  & 2.54\\
\hline
\end{tabular}
\caption{\label{coh_sys} Contributions to the systematic errors in the coherent fraction.  Errors are given in percent coherent.}
\end{table}

As a test of the model dependence of the coherent fraction, the data were refit after significant modifications of the MC model parameters.  Five variations on the NUANCE model were explored:
\begin{table}[b]
\begin{tabular}{@{\extracolsep{1.5cm}}lcrr} \hline
\tablehead{1}{l}{b}{Variation} & 
\tablehead{1}{c}{b}{Coherent}  & 
\tablehead{2}{c}{b}{Avg. C.L. (\%)} \\ 
 & \tablehead{1}{c}{b}{Fraction (\%)} & 
\tablehead{1}{c}{b}{Coh.} & 
\tablehead{1}{c}{b}{No Coh.} \\ \hline
Default Model            & 19.5$\pm$1.1    &  5.97  & 1.8$\times10^{-16}$ \\
M$_A$ Coherent           & 19.1$\pm$1.1    &  5.73  & 3.8$\times10^{-17}$ \\
No Diffractive           & 17.9$\pm$1.0    & 11.63  & 5.7$\times10^{-17}$ \\
M$_A^{1\pi}$ Resonant Hi & 17.9$\pm$1.1    &  3.27  & 1.5$\times10^{-14}$ \\
M$_A^{1\pi}$ Resonant Lo & 21.1$\pm$1.1    &  5.00  & 7.2$\times10^{-22}$ \\
Binding Energy Hi        & 19.4$\pm$1.1    &  5.85  & 6.4$\times10^{-16}$ \\
Binding Energy Lo        & 19.6$\pm$1.1    &  6.29  & 2.2$\times10^{-17}$ \\
Fermi Momentum Hi        & 18.2$\pm$1.1    &  3.29  & 1.3$\times10^{-15}$ \\
Fermi Momentum Lo        & 21.0$\pm$1.1    &  4.24  & 4.6$\times10^{-23}$ \\
Isotropic $\Delta$ Decay & 18.1$\pm$1.1    &  1.88  & 1.8$\times10^{-16}$ \\
Pure Spin 3/2            & 20.8$\pm$1.0    &  3.49  & 1.7$\times10^{-18}$ \\ 
Pure Spin 1/2            & 16.9$\pm$1.2    &  0.01  & 1.3$\times10^{-19}$ \\ 
\hline
\end{tabular}
\vspace{-0.2in}
\caption{\label{coh_model_study} Average fitted coherent fractions and confidence levels for fits with several variations of the cross section model, including the study of the angular distribution of the $\Delta$ decay.  The default model's angular distribution is given by the model of Rein and Sehgal~\cite{Rein:1982pf}.  The confidence level is also given for fits where the coherent fraction is fixed to zero.}
\end{table}
\begin{itemize}
\item{The coherent axial mass assumed in the Rein-Sehgal coherent model \cite{Rein:1982pf} was decreased by a factor of three. This particular excursion was chosen because it is what would be needed to bring the predicted coherent $\pi^0$ cross section into agreement with the measured MiniBooNE coherent rate.  While the normalization is given by the fit, this change does alter the distribution of kinematics for the coherent events, and therefore can change the fit result.}
\item{Diffractive events, which arise from coherent scattering off hydrogen targets, account for $16\%$ of all coherent events in MiniBooNE.  In this variation diffractive events were removed from the coherent fit template.  The diffractive contribution tends  to be slightly less forward peaked than the coherent scatters off carbon, and hence their removal impacts the shape of the coherent fit template.}
\item{The axial mass, $M_A^{1\pi}$, assumed in the Rein-Sehgal resonant model \cite{Rein:1980wg} was varied by $\pm 25\%$ of its default value.  Altering the axial form factor parameter affects the resonant contribution at low $Q^2$.}
\item{The nuclear model affecting resonant $\pi^0$ events was modified:
\begin{itemize}
\item{The binding energy in the Fermi Gas model was varied by 177\% from 34~MeV to 60~MeV.} 
\item{The Fermi momentum in the Fermi Gas model was varied by 72\% from 246~MeV/c to 423~MeV/c.}
\end{itemize}
\noindent
The variations are quite large and specifically impact the resonant predictions at low $Q^2$. By reducing the nuclear effects, one can test whether large changes to the Fermi Gas model, not from coherent scattering, can improve agreement to data in the most forward (or low $Q^2$) region.}
\item{The $\Delta$ decay angular distribution was varied from the default model, which is from Rein and Sehgal~\cite{Rein:1982pf}.  The variations included isotropic decay in the center of mass frame, a pure spin 1/2 decay, and a pure spin 3/2 decay.  The default model in NUANCE is isotropic decay and, in making the switch to the Rein and Sehgal decay distribution, a significant shift in the coherent fraction was observed.}
\end{itemize}
\noindent
The fitted coherent fraction and fit confidence levels for each of these studies are given in Table~\ref{coh_model_study}.  In addition, the fit confidence level for fits with the coherent fraction fixed to zero are also given.  The zero coherent fits show clearly show that none of these variations prefer a coherent free production model as has been suggested by the K2K charged pion study~\cite{Hasegawa:2005td}.  

\section{Conclusions}
Neutral current $\pi^0$ production is both a major potential background to the MiniBooNE oscillation analysis and an opportunity to make high impact measurements in neutrino cross sections (with the world's largest data set of 0.5 to 2~GeV neutrino interactions).  The analysis described here has resulted in a direct measurement of $\pi^0$ production in the MiniBooNE detector, which is critically important for estimating the $\pi^0$ misidentification background to $\nu_e$ appearance.  In addition, the coherent $\pi^0$ production, relative to total exclusive $\pi^0$ production, was measured and found to be (19.5 $\pm$1.1 (stat) $\pm$2.5 (sys))\%.


\begin{theacknowledgments}
The author aknowledges the kind support of the Thomas F. and Kate Miller Jeffress Memorial Trust.
\end{theacknowledgments}

\bibliographystyle{aipproc}   
\bibliography{nuint07}
\end{document}